\documentclass[12pt,a4paper]{article}
\usepackage{a4wide}
\usepackage{graphicx}
\usepackage{amsmath}
\usepackage{feynmp-auto}
\usepackage{subfig}

\begin{document}

\title{\bf{Lepton Dipole Moments}\\\small{Physics In Collision 2015}}
\date{}
\maketitle
\bigskip\bigskip


\begin{raggedright}  
\vspace{-30pt}
{\it Adam West\index{West, A. D.}\\
Department of Physics\\
Yale University\\
217 Prospect Street\\
New Haven, CT, USA}
\bigskip\bigskip
\end{raggedright}

\section{Introduction}
The Standard Model (SM) of physics provides our best description of the world, capable of making stunningly accurate predictions, but it is conspicuously incomplete. The continued testing of and search for extensions to the SM currently represents much of the focus of modern particle physics. Evidence for new physics (NP) beyond the SM is expected to reveal itself through new types of interactions. As such, the electromagnetic moments of fundamental particles are excellent probes for NP, which would be evidenced by a deviation of these moments from their predicted values.

The electric and magnetic dipole moments (EDMs and MDMs) of a fundamental particle are given by
\begin{equation}
\vec{\mu}=\mu\hat{S}=g\frac{q\hbar}{2mc}\hat{S},\quad{\rm and}\quad\vec{d}=d\hat{S}=\eta\frac{q\hbar}{2mc}\hat{S}
\end{equation}
respectively, where $g$ is the unitless $g$-factor, $q$ the charge, $m$ the mass. $d$ or $\eta$ describe the size of the EDM. CGS units are used unless specified. $g=2$ to zeroth order and QED provides perturbative corrections at successively higher orders in $\alpha$. By contrast $d$ is zero to zeroth order since a non-zero value requires $CP$-violation. This can be intuitively understood by considering $\vec{d}_{\ell}$ aligned with the spin axis $\hat{S}$. $P$-reversal reverses the direction of $\vec{d}_{\ell}$ but not $\hat{S}$; $T$-reversal reverses $\hat{S}$ but not $\vec{d}_{\ell}$. Assuming the $CPT$ theorem the intrinsic $T$-violation can be recast as $CP$-violation. The SM predictions for EDMs rely on four-loop processes, predicting values far below the reach of current experiments. For example $d^{\rm SM}_e\sim10^{-38}~e\cdot{\rm cm}$ \cite{Pospelov2014}.

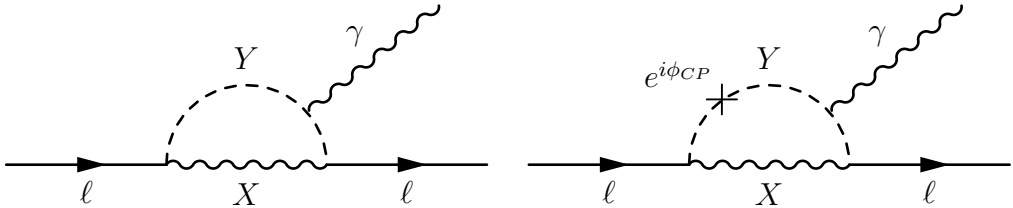
\begin{figure}[!ht]
\centering
\begin{fmffile}{susy_MDM}
\begin{fmfgraph*}(180,120)
\fmfleft{i1}
\fmfright{o1,o2,o3}
\fmftop{o6}
\fmf{fermion,tension=1,label=$\ell$,l.side=right}{i1,v1}
\fmf{wiggly,label=$X$,l.side=right,l.dist=7}{v1,v2}
\fmf{fermion,tension=1,label=$\ell$,l.side=right}{v2,o2}
\fmf{dashes,left,tension=0,label=$Y$,l.side=left}{v1,v2}
\fmffreeze
\fmf{wiggly,label=$\gamma$,l.side=left}{o6,o3}
\fmfforce{0.63w,0.67h}{o6}
\fmfforce{0.5w,0.5h}{o1}
\end{fmfgraph*}
\end{fmffile}
\begin{fmffile}{susy_EDM}
\begin{fmfgraph*}(180,120)
\fmfleft{i1}
\fmfright{o1,o2,o3}
\fmftop{o6}
\fmf{fermion,tension=1,label=$\ell$,l.side=right}{i1,v1}
\fmf{wiggly,label=$X$,l.side=right,l.dist=7}{v1,v2}
\fmf{fermion,tension=1,label=$\ell$,l.side=right}{v2,o2}
\fmf{dashes,left,tension=0,label=$Y$,l.side=left}{v1,v2}
\fmffreeze
\fmf{wiggly,label=$\gamma$,l.side=left}{o6,o3}
\fmfforce{0.63w,0.67h}{o6}
\fmfforce{0.5w,0.5h}{o1}
\fmfiv{lab=\rotatebox[origin=c]{45}{\scalebox{1.5}{$\times$}},label.dist=-1.}{(0.45w,0.636h)}
\fmfiv{lab=$e^{i\phi_{CP}}$,label.dist=-1.}{(0.38w,0.75h)}
\end{fmfgraph*}
\end{fmffile}
\vspace{-48pt}
\caption{Left: (Right:) Example Feynman diagram of a NP contribution to a magnetic (electric) dipole moment  \cite{Graesser2002}. $\ell$ represents a lepton and $\gamma$ a photon, while $X$ and $Y$ are some new particles. Note that the loop contains a $CP$-violating phase for contributions to an electric dipole moment.}
\label{fig:SUSY_fds}
\end{figure}

Many NP models give significant modification to the predicted dipole moments, often via interactions at the one-loop level. An example of such an interaction is shown in Fig.~\ref{fig:SUSY_fds}.
One immediately sees that the interactions are very closely analogous for MDMs and EDMs. The primary difference is the presence of a $CP$-violating phase in the latter. An observed $d_{\ell}$ above the predicted SM level would indicate additional $CP$-violation beyond the SM that could help explain the observed baryon asymmetry of the universe \cite{Sakharov1967}. 

For interactions such as those in Fig.~\ref{fig:SUSY_fds} we can write (in natural units)
\begin{equation}
\mu_{\ell}^{\rm NP}\sim e\left(\frac{k_{\mu}^2}{4\pi}\right)^n\frac{m_{\ell}}{\Lambda^2_{\mu}}\quad\quad
d_{\ell}^{\rm NP}\sim e\left(\frac{k_d^2}{4\pi}\right)^n\frac{m_{\ell}}{\Lambda^2_d}\sin\phi_{CP},
\label{eq:mass_scales}
\end{equation}
where $\mu_{\ell}^{\rm NP}$ and $d_{\ell}^{\rm NP}$ are the MDM and EDM contributions from NP. $\Lambda$ is the NP mass scale, $k$ is the coupling strength between lepton $\ell$ and new particle, $n$ is the loop order, $\phi_{CP}$ is a $CP$-violating phase and $m_{\ell}$ is the lepton mass. Because many beyond-SM (BSM) theories (including variants of supersymmetry (SUSY)) have flavour universality built in, there is a `na\"{i}ve scaling' between flavours exhibited in Equation~\ref{eq:mass_scales}:
\begin{equation}
\frac{\mu^{\rm NP}_{\ell_i}}{\mu^{\rm NP}_{\ell_j}}\propto\frac{d^{\rm NP}_{\ell_i}}{d^{\rm NP}_{\ell_j}}\propto\frac{m_{\ell_i}}{m_{\ell_j}}.
\label{eq:NS1}
\end{equation}
The resulting complementarity of lepton flavours enhances the measurement of their dipole moments as a tool for searching for NP by allowing tests of flavour-universality.

The most precise determinations of lepton dipole moments are performed as measurements of spin-precession frequency, a technique favouring particles with longer lifetimes. In such experiments an applied magnetic or electric field exerts a torque on the dipole moment, and hence on the spin, causing it to precess about the applied field. The corresponding statistical sensitivity, limited by quantum projection noise, can be written as
\begin{equation}
\label{eq:statsens}
\delta\mu=\frac{\hbar}{2\mathcal{C}\tau B\sqrt{\dot{N}T}},\quad\quad\delta d=\frac{\hbar}{2\mathcal{C}\tau\mathcal{E}_{\rm eff}\sqrt{\dot{N}T}},
\end{equation}
where $\mathcal{C}$ is the contrast (a measure of how well a spin state can be resolved), $\tau$ the coherence time, $B$ the magnetic field acting on a MDM, $\mathcal{E}_{\rm eff}$ the effective electric field acting on an EDM, and $\dot{N}T$ is the total number of measurements.

In the rest of this paper I will describe the current progress of these measurements, discussing briefly their impact as tests of the SM, probes of NP, and the relationship between these measurements and those at colliders. For more details the interested reader is directed to the excellent review provided by \cite{Roberts2009}.

\section{Magnetic Dipole Moments}
Measurement of magnetic dipole moments is usually described in terms of the quantity $a$:
\begin{equation}
a_{\ell}=(g_{\ell}-2)/2.
\end{equation}
This represents the deviation from the `Dirac' magnetic moment. Experiments to measure $a$ generally search for deviations from the SM prediction, $\Delta a=a-a^{\rm SM}$. Under na\"{i}ve scaling (see Eq.~\ref{eq:NS1}) we then have $\Delta a_{\ell}\sim m_{\ell}^2/\Lambda_{\ell}^2$. This notation shall be used from here on.

\subsection{Electron}
The small mass of the electron makes it less sensitive than the muon and tau to many types of NP (cf. Equation~\ref{eq:mass_scales}), however it can be used to search for new particles with low mass and weak couplings. The electron is also the most stable lepton, leading to it having the most precisely measured value of $a$. This measurement represents the most precise test of the SM.

\begin{figure}[h]
\centering
\includegraphics[width=12cm]{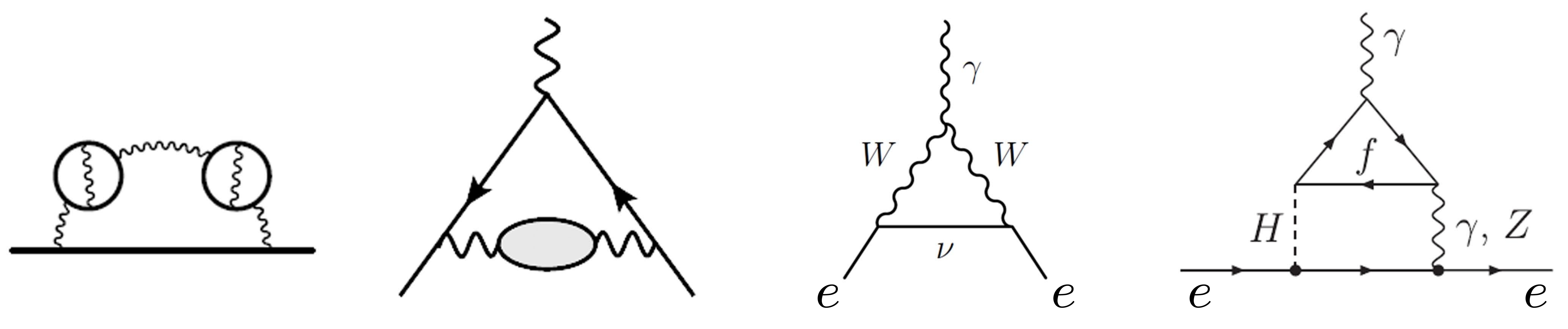}
\caption{A selection of the Feynman diagrams contributing to the calculation of $a_e$. They correspond to, from left to right, QED, hadronic, electroweak and loop Higgs contributions. The grey circle indicates loops with hadronic particles, e.g. pions. Reprinted with permission from \cite{Aoyama2012,Kurz2014,Czarnecki2003,Gnendiger2013}.}
\label{fig:e_g-2_fds}
\end{figure}

State-of-the-art calculations currently predict $a_e$ with 13 digits of precision \cite{Aoyama2015}:
\begin{equation}
\label{eq:aetheory}
a_e{\rm (theory)}=1~159~652~181.643~(25)(23)(16)(763)\times10^{-12}.
\end{equation}
This calculation draws from all areas of the Standard Model; Fig.~\ref{fig:e_g-2_fds} shows a selection of the contributing Feynman diagrams.
We can decompose $a_e$ into these various contributions and write
\begin{equation}
a_e=a_e^{\rm QED}+a_e^{\rm Hadronic}+a_e^{\rm Electroweak}+\cdots,
\label{eq:a_comp}
\end{equation}
grouping the contributions to $a_e$ by the type of interaction involved. The first three uncertainties in Eq.~\ref{eq:aetheory} correspond to 4-loop QED, 5-loop QED, and hadronic/electroweak contributions. The final uncertainty is from that on the value of the fine structure constant $\alpha$, as determined from atomic photon recoil measurements \cite{Bouchendira2011}.

\begin{figure}[!ht]
\centering
\includegraphics[width=6cm]{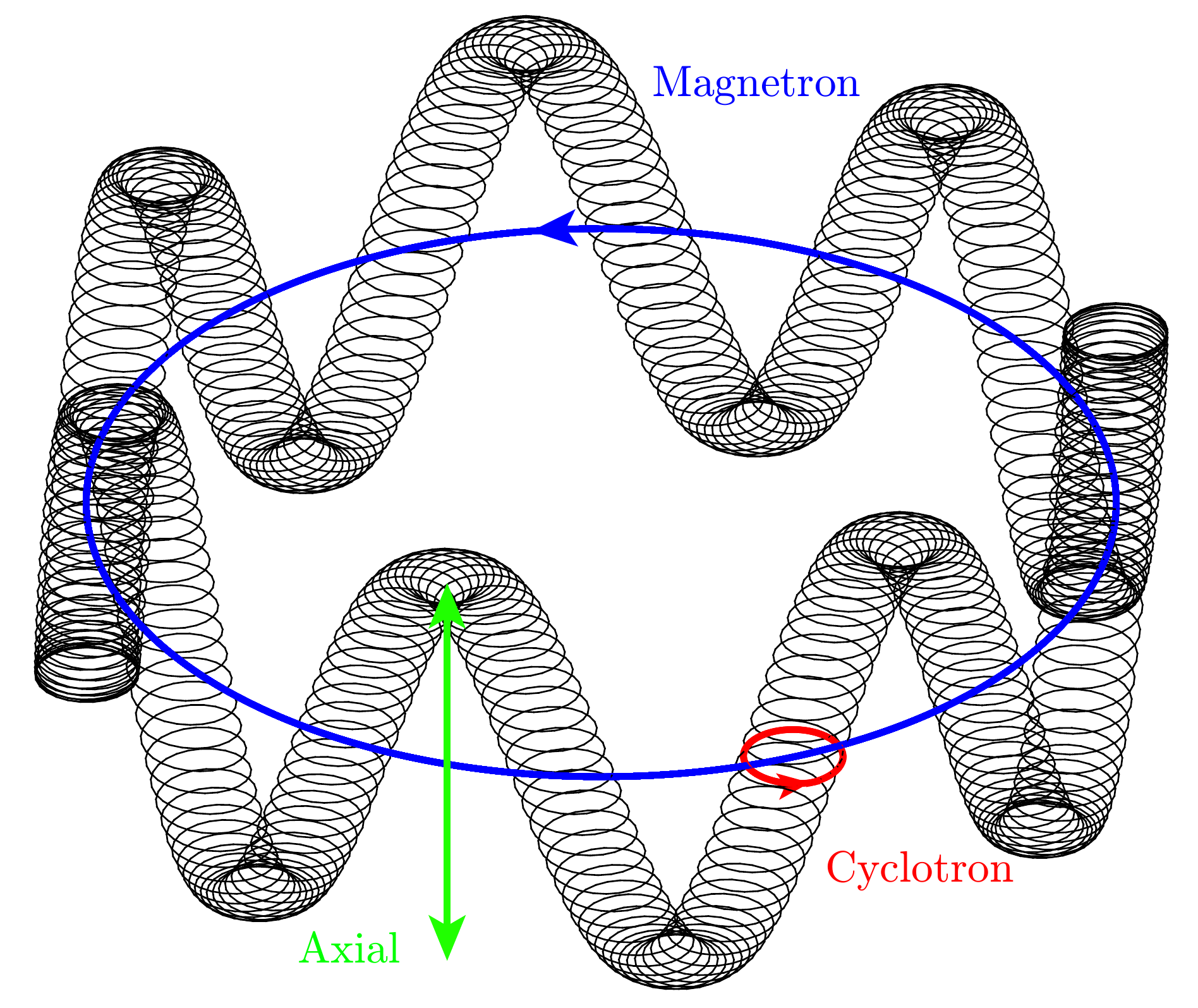}\hspace{40pt}
\includegraphics[width=6cm]{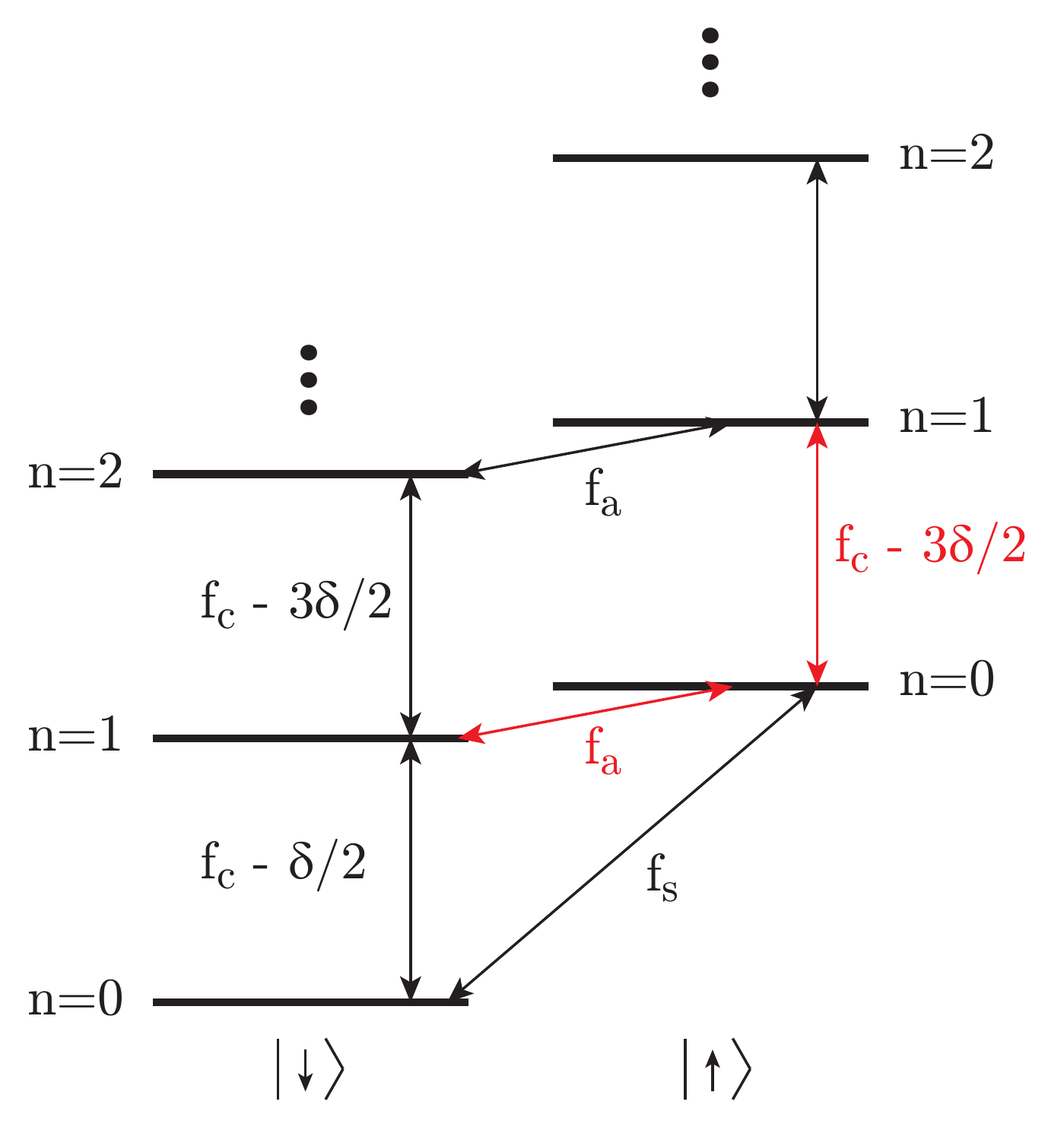}
\caption{Left: Electron motion inside a Penning trap. Right: The corresponding ladder of cyclotron energy states for both spin states. See main text for description.}
\label{fig:penning}
\end{figure}

The most precise measurement of $a_e$ is currently that of the Gabrielse group at Harvard \cite{Hanneke2008}, achieved using quantum jump spectroscopy in a Penning trap. Such a trap consists, in its simplest form, of a superposition of a uniform magnetic field and a quadrupolar electric field \cite{Brown1986}. The resulting motion of the electron is illustrated on the left-hand side of Fig.~\ref{fig:penning}, consisting of cyclotron (axial) motion induced by the magnetic (electric) field, and a slower magnetron precession from the combined fields. A `magnetic bottle' couples the cyclotron and spin states to the axial motion which is detected via the image current induced in the trapping electrodes by the electron's motion.

The corresponding energy spectrum is shown on the right-hand side of Fig.~\ref{fig:penning} --- there are two ladders of cyclotron states (quantum number $n$) for the spin up and down electron states\footnote{Splitting into axial and magnetron levels is not shown.}. $f_c=qB/(2\pi mc)$ is the cyclotron frequency, $f_s=(g/2)qB/(2\pi mc)$ the spin-flip frequency, and $\delta$ is a small relativistic modification. For $g=2$ we have $f_c=f_s$. Thus, a measurement of the anomalous frequency $f_a=f_s-f_c+\delta/2$ provides a direct measure of $a_e$:
\begin{equation}
a_e^{\rm (expt)}=(f_a-\delta/2)\frac{2\pi mc}{qB}=1~159~652~180.73~(28)\times10^{-12}.
\end{equation}
For more details on the experimental procedure the interested reader is directed to \cite{Dorr2013,Hanneke2007}.

The experimental and theoretical values are in very good agreement when $\alpha$ is taken from an independent experiment whose result does not rely on $a_e$. Conversely, by assuming the validity of the SM one can use the measured value of $a_e$ to deduce $\alpha$. This yields a result with an order of magnitude greater precision than is possible via other methods \cite{Gabrielse2009}:
\begin{equation}
\alpha^{-1}=137.035~999~084~(51).
\end{equation}

The comparison of the measured and calculated values of $a_e$ does not substantially probe NP at high energy scales. It is anticipated that NP contributions to $a_e$ will be around the size of the electroweak contribution ($0.02973~(52)\times10^{-12}$ \cite{Czarnecki1996,Czarnecki1995}), which is smaller than the overall uncertainty by an order of magnitude. Further improvement in the precision of the experimental determination of $a_e$, and in the independent determination of $\alpha$, would significantly expand the NP effects probed. However, at the current precision the comparison of $a_e^{\rm expt}$ and $a_e^{\rm theory}$ makes it possible to set limits on the existence of `dark photons', i.e. light vector bosons that couple feebly to the electron \cite{Endo2012}.

\subsection{Muon}
\label{sec:mugm2}
The current best experimental measure of $a_{\mu}$ (average of $a_{\mu^{\pm}}$) is provided by the E821 experiment at Brookhaven National Lab \cite{Bennett2006,Mohr2008}:
\begin{equation}
a_{\mu}^{\rm expt}=116~592~089~(63)\times10^{-11}.
\end{equation}

\begin{figure}[!ht]
\centering
\includegraphics[width=8cm]{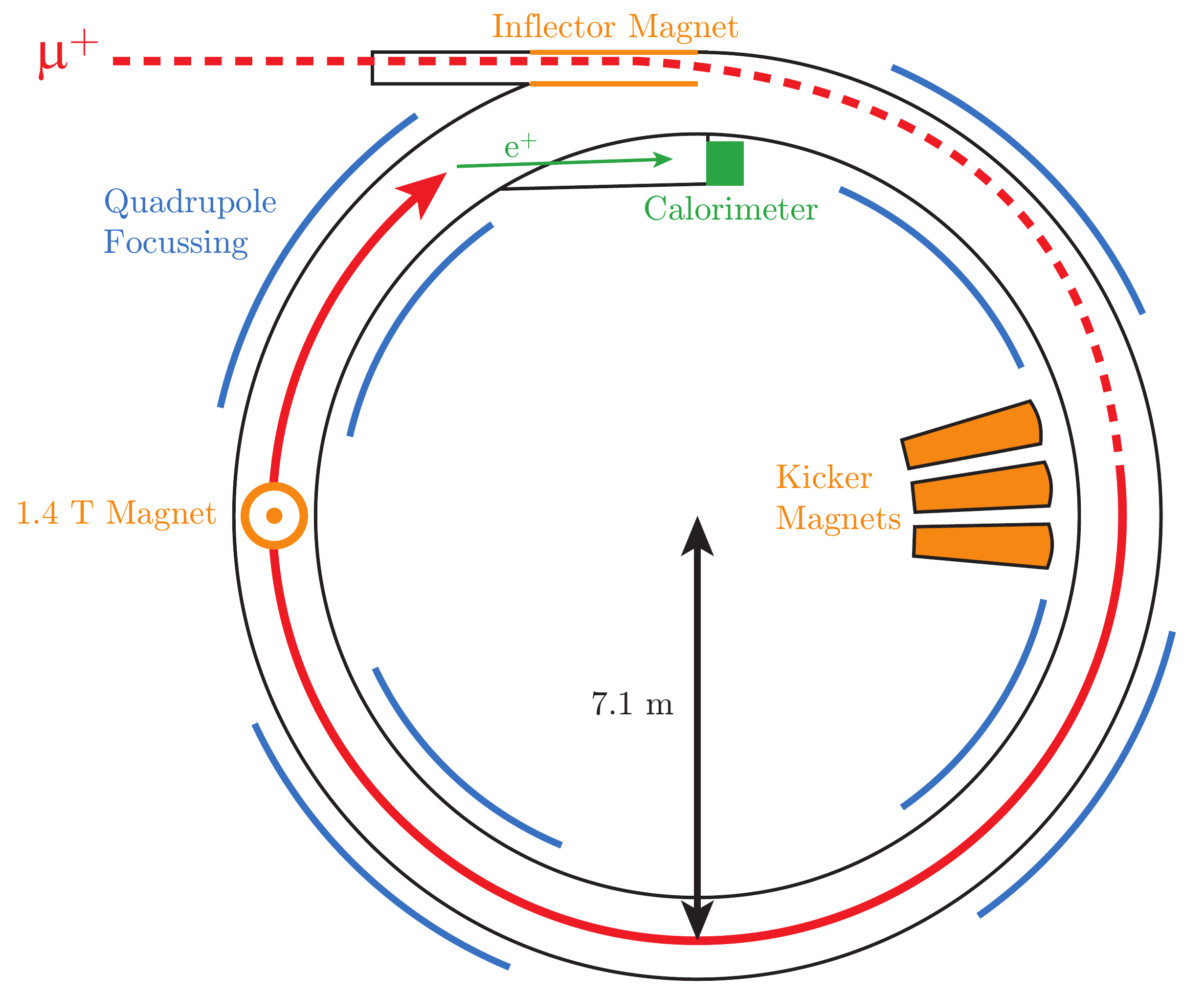}
\includegraphics[width=8cm]{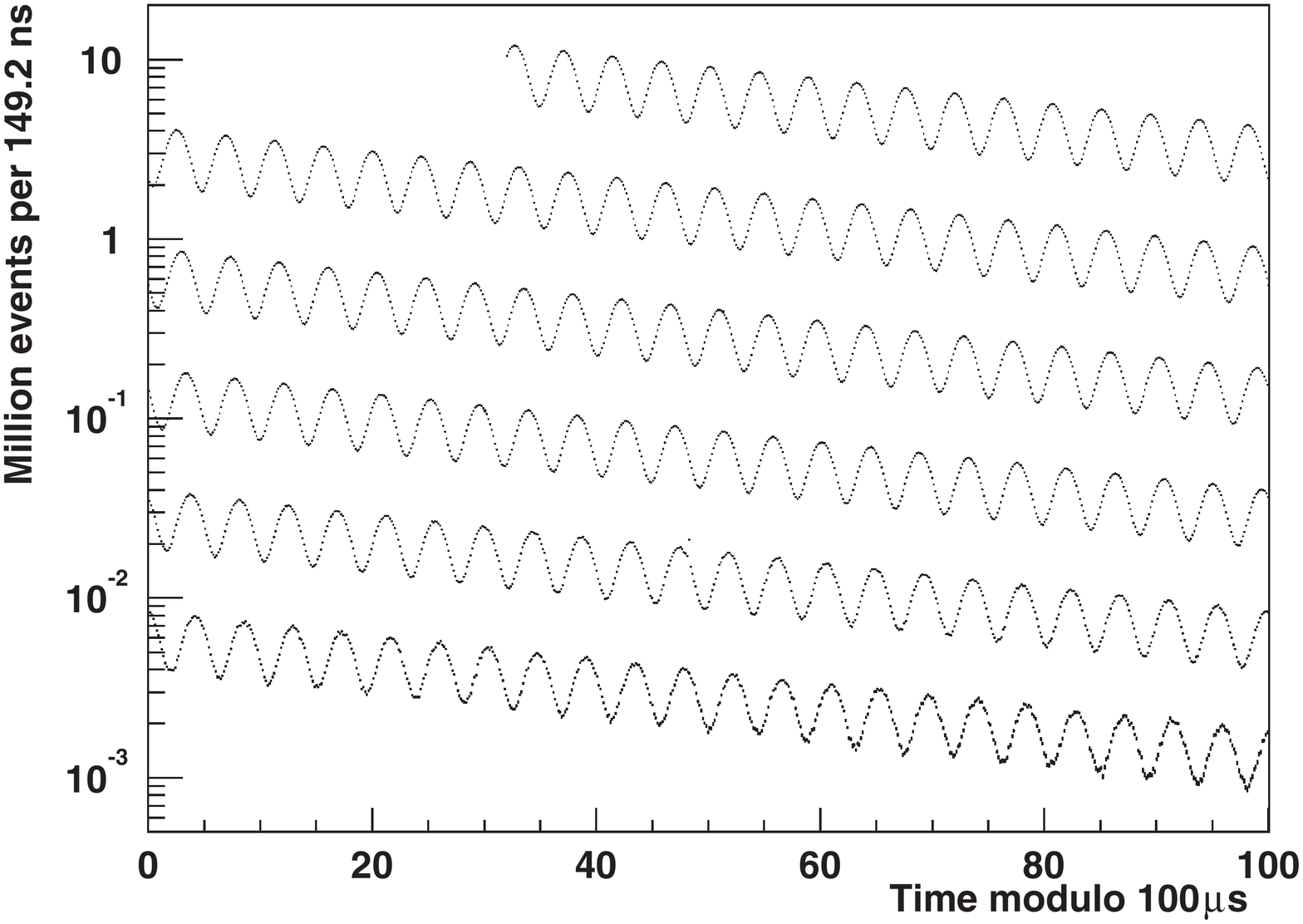}
\caption{Left: Schematic of the E821 experiment. Right: Signal from E821. Data are shown modulo 100~$\mu$s. Reprinted with permission from \cite{Bennett2006}. See main text for description.}
\label{fig:mugm2}
\end{figure}

Experimental measurement of the anomalous magnetic moment is substantially more difficult for the muon, owing to its production method and short lifetime (2.2~$\mu$s). However, the measurement technique is entirely analogous: one searches for a discrepancy between the frequencies of cyclotron motion and spin precession. This is performed in a storage ring, shown schematically on the left-hand side of Fig.~\ref{fig:mugm2}.

A proton beam incident on a target produces a large number of pions which subsequently decay into muons. The selected muon momentum gives a relativistically enhanced lifetime of around $64~\mu$s. The muon beam is injected at a velocity $\vec{v}$ into a 7.1~m radius ring where there is a 1.4~T vertical magnetic field, $\vec{B}$, which produces cyclotron motion matching the ring radius. Electrostatic focussing of the beam is provided by a series of quadrupole lenses around the ring.

The associated anomalous frequency can be written as
\begin{equation}
\label{eq:mu_wa}
\vec{\omega_a}=\frac{e}{mc}\left[a_{\mu}\vec{B}-\left(a_{\mu}-\frac{1}{\gamma^2-1}\right)\vec{\beta}\times\vec{E}\right]\approx a_{\mu}\frac{e\vec{B}}{mc},
\end{equation}
where $e$ is the elementary charge, $m$ is the mass, $c$ is the speed of light, $\gamma$ is the Lorentz factor, $\vec{\beta}=\vec{v}/c$ and $\vec{E}$ is the electric field present due to the electrostatic focussing. The term associated with the electric field is approximately eliminated by choosing a `magic' momentum where $\gamma=29.3$.

The muon decays to an electron and a pair of neutrinos, with high-energy electrons preferentially emitted parallel (anti-parallel) to the anti-muon (muon) spin. The anomalous precession modulates the electron flux into the detection regions around the ring, consisting of scintillators and photomultipliers. This yields the signal shown on the right-hand side of Fig.~\ref{fig:mugm2} \cite{Bennett2006}. The exponential decrease due to muon decay and the sinusoidal oscillation due to the anomalous precession are clearly seen. Careful measurement of the magnetic field experienced by the muons allows $a_{\mu}$ to be deduced from the oscillation frequency.

Two independent analyses of the hadronic contributions to $a_{\mu}$, which rely on cross-section data from colliders, lead to consistent predictions of its value in the SM:
\begin{align}
a_{\mu}^{\rm theory}=116~591~802~(42)(26)(2)&\times10^{-11}~\cite{Davier2012},\\
116~591~828~(43)(26)(2)&\times10^{-11}~\cite{Hagiwara2011}.
\end{align}
The quoted error bars correspond to leading order hadronic contributions, higher order hadronic contributions and other contributions respectively. For either theoretical value we find that $\Delta a_{\mu}=a_{\mu}^{\rm expt}-a_{\mu}^{\rm theory}$ differs from zero by more than $3\sigma$. This could be evidence for NP.

\begin{figure}[!ht]
\centering
\includegraphics[width=15cm]{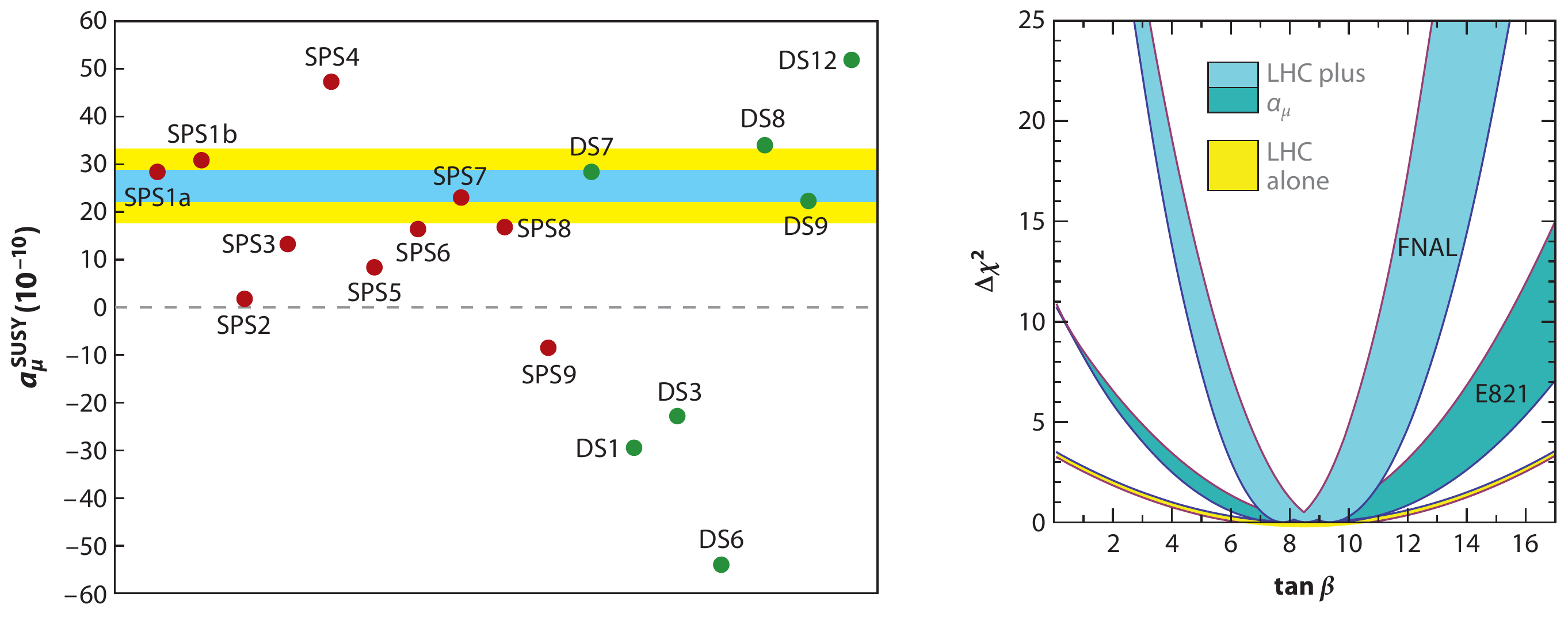}
\caption{Left: Predictions of the supersymmetric contribution to $a_{\mu}$ from various models. Right: $\Delta\chi^2$ parabolas for the $\tan\beta$ parameter showing the complementarity of the LHC and the $a_{\mu}$ measurements, reprinted with permission from \cite{Miller2012}. See main text for description.}
\label{fig:mugm2_BSM}
\end{figure}

Due to its significantly higher mass, the electroweak and hadronic contributions to $a_{\mu}$ are more significant than in the electron. Correspondingly, $a_{\mu}$ is around 40,000 times more sensitive to NP at the electroweak scale and is already helping constrain some BSM theories in a manner complementary to the LHC. For example, the left-hand plot of Fig.~\ref{fig:mugm2_BSM} shows a series of predictions for the SUSY contribution to $a_{\mu}$ \cite{Miller2012}. The horizontal bands indicate the region constrained by the $a_{\mu}$ measurement. Assuming the $a_{\mu}$ discrepancy is due to NP, the E821 measurement discriminates well between these theories. This is particularly useful for the `DS' points --- degenerate solutions among which the LHC cannot discriminate \cite{Adam2011}.
Within the constraints of a theory such as SPS1a\footnote{The SPS points are the `Snowmass Points and Slopes' --- accepted benchmarks for minimal supersymmetry (MSSM) \cite{Allanach2002}.}, the measurement of $a_{\mu}$ can help to constrain $\tan\beta$ (ratio of Higgs vacuum expectation values). This is illustrated in the right-hand plot of Fig.~\ref{fig:mugm2_BSM} which shows the corresponding $\Delta\chi^2=\left[\left(a_{\mu}^{\rm SUSY}(\tan\beta)-\Delta a_{\mu}\right)/\delta a_{\mu}\right]^2$ parabolas, where $\delta a_{\mu}$ is the precision of the measurement of $a_{\mu}$. The yellow band is that derived from the LHC. The darker blue band also includes the measurement of $a_{\mu}$. The lighter blue band includes the expected sensitivity improvement from the future measurement of $a_{\mu}$ at FNAL. As can be seen, the two experiments would be complementary in constraining $\tan\beta$.

In general we can write a SUSY contribution to $a_{\mu}$ as \cite{Miller2012}
\begin{equation}
a_{\mu}^{\rm SUSY}\approx 13\times10^{-10}\left(\frac{100~{\rm GeV}}{M_{\rm SUSY}}\right)^2\tan\beta~{\rm sgn}(\mu),
\end{equation}
where $M_{\rm SUSY}$ is the superpartner mass scale and here ${\rm sgn}(\mu)$ is the sign of the Higgsino mass parameter. Assuming $\tan\beta\sim10~(100)$ we see that $M_{\rm SUSY}$ must be around 200~GeV (700~GeV) to explain the $a_{\mu}$ discrepancy. This presents a source of tension between LHC data and the $a_{\mu}$ measurement, which could be resolved by a reduction in $\Delta a_{\mu}$, fine tuning of SUSY models, or alternative BSM theories.

With the presence of $\Delta a_{\mu}$ it is interesting to consider the complementarity with other lepton moments. For many flavour-universal theories na\"{i}ve scaling allows one to write \cite{Giudice2012}
\begin{equation}
\label{eq:et_gm2_disc}
\Delta a^{\rm NS}_e=\frac{m_e^2}{m_{\mu}^2}~\Delta a^{\rm NS}_{\mu}=6\times10^{-14},\quad\quad\Delta a_{\tau}=\frac{m_{\tau}^2}{m_{\mu}^2}\Delta a_{\mu}=7\times10^{-7}.
\end{equation}
The current precision of $a_e^{\rm expt}$ is close to the discrepancy from SM predicted by na\"{i}ve scaling --- another reason why further improvement in the precision of $a_e^{\rm expt}$ (and $\alpha$, as needed to improve $a_e^{\rm theory}$) is of interest. Conversely, the precision of $a_{\tau}$ is still quite far from the above predicted discrepancy. Of course, one should bear in mind that there are theories for which such na\"{i}ve scaling is not applicable.

\subsection{Tau}
\begin{figure}
\centering
\subfloat
{
\raisebox{25pt}
{
\begin{fmffile}{diagram}
\begin{fmfgraph*}(160,110)
\fmfleft{i1,i2}
\fmfright{o1,o2,o3,o4}
\fmf{fermion,label=$e^+$}{o1,v1,i1}
\fmf{fermion,label=$e^-$,label.side=left}{i2,v2,o4}
\fmffreeze
\fmf{photon}{v1,v3}
\fmf{photon}{v2,v4}
\fmf{fermion,tension=0,label=$\tau^+$}{o2,v3}
\fmf{fermion,tension=0,label=$\tau^-$}{v4,o3}
\fmf{plain,tension=0.92}{v3,v4}
\end{fmfgraph*}
\end{fmffile}
}
}
\subfloat{\includegraphics[width=5cm]{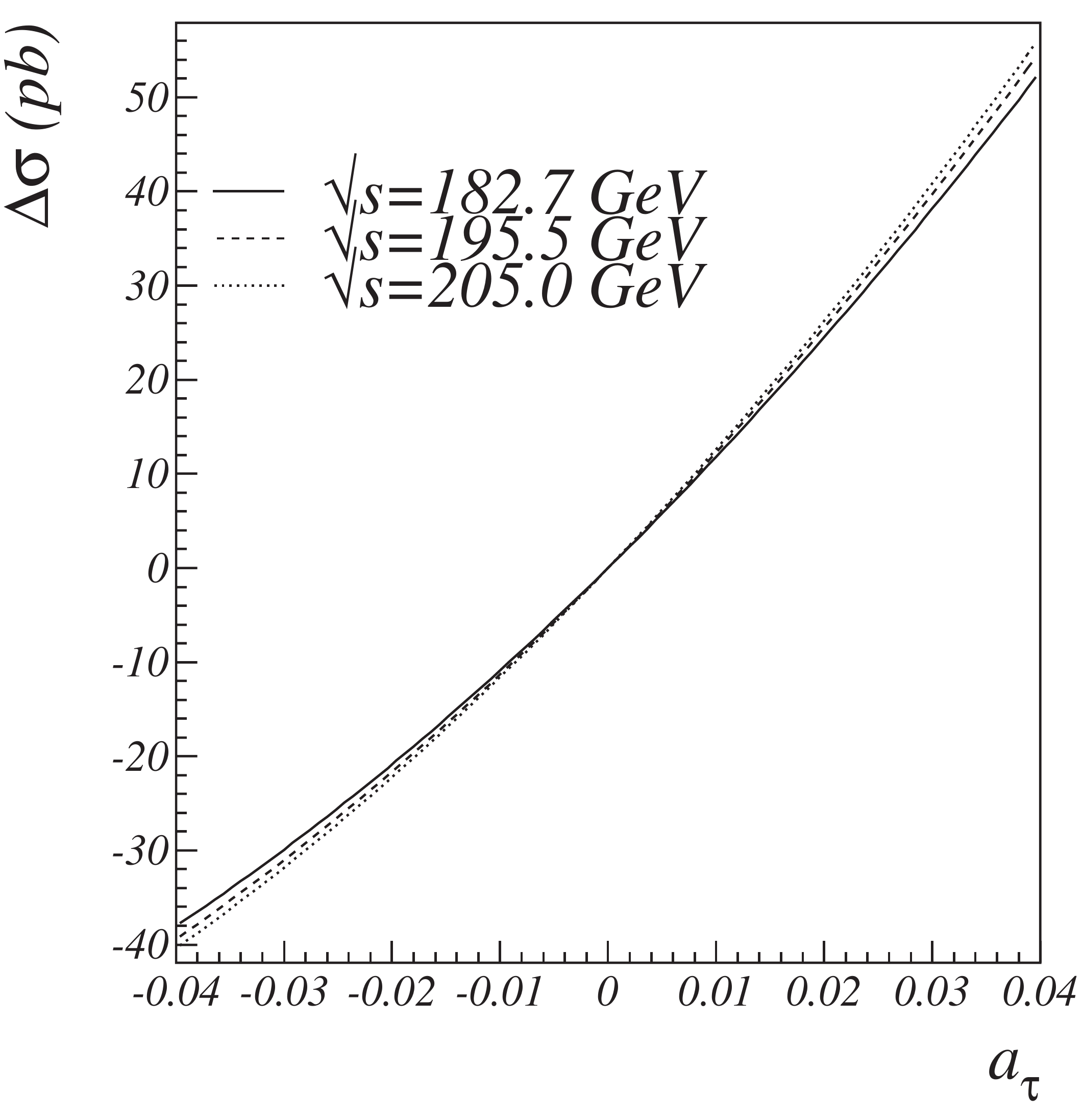}}
\caption{Left: $e^-e^+\tau^-\tau^+$ production from $e^-e^+$ via a virtual photon pair. Right: Calculated dependence of the $e^-e^+\rightarrow e^-e^+\tau^-\tau^+$ cross section on $a_{\tau}$. Reprinted with permission from \cite{DELPHI2004}.}
\label{fig:taugm2}
\end{figure}

Because of its higher mass, the tau magnetic anomaly, $a_{\tau}$, is expected to be around 300 times more sensitive to NP than that of the muon (assuming na\"{i}ve scaling). However the intrinsic difficulty of producing and observing interactions of the tau has so far precluded precision measurements of its dipole moments; its 0.3~ps (rest frame) lifetime prevents a frequency measurement. The current limit on $d_{\tau}$ was inferred by the DELPHI collaboration from their measurement of the cross-section for the process shown in the left-hand plot of Fig.~\ref{fig:taugm2} \cite{DELPHI2004} --- electron-positron collisons produce $\tau$ pairs via two virtual photons. The $\gamma\tau\tau$ vertices of this process are sensitive to both $a_{\tau}$ and $d_{\tau}$.
The calculated cross-section for this process is shown in the right-hand plot of Fig.~\ref{fig:taugm2} for a range of values of $a_{\tau}$ and collision energy $\sqrt{s}$, showing how a precision measurement of the cross-section directly translates into a measurement of $a_{\tau}$, whilst assuming the validity of the QED theory used in the calculation.

The measured cross-section of $429\pm17$~pb agrees with the SM prediction of $447.7\pm0.3$~pb at the level of $1\sigma$. This provides a test of QED at fourth order in $\alpha$ and a limit for $a_{\tau}$ of
\begin{equation}
-0.052<a_{\tau}<0.013~(95\%~{\rm C.L.}).
\end{equation}
The relative imprecision of this limit compared to measurements of the muon and electron means it is a less effective probe of NP. There are a number of proposals for measuring $a_{\tau}$ with an order of magnitude greater precision, for example by studying the radiative decay $W\rightarrow\tau\nu\gamma$ \cite{Samuel1994}, $\tau$ production from heavy ions \cite{Aguila1991} or even a direct measurement of spin precession using a strong motional magnetic field \cite{Chen1992}.

\subsection{Neutrinos}
Within the SM, a measurement of a dipole moment of a neutrino would be of critical importance in showing it to be a Dirac particle \cite{Winter2008}. While the SM predicts Dirac neutrinos to have a magnetic moment $\mu_{\nu}<10^{-19}~\mu_{\rm B}$, some BSM theories predict significantly larger dipole moments ($\sim10^{-14}$--$10^{-10}\mu_{\rm B}$) \cite{Aboubrahim2014}. The fact that these moments may be detectable by future experiments illustrates one way in which the neutrino can probe NP.

\begin{figure}[!ht]
\centering
\includegraphics[width=7cm]{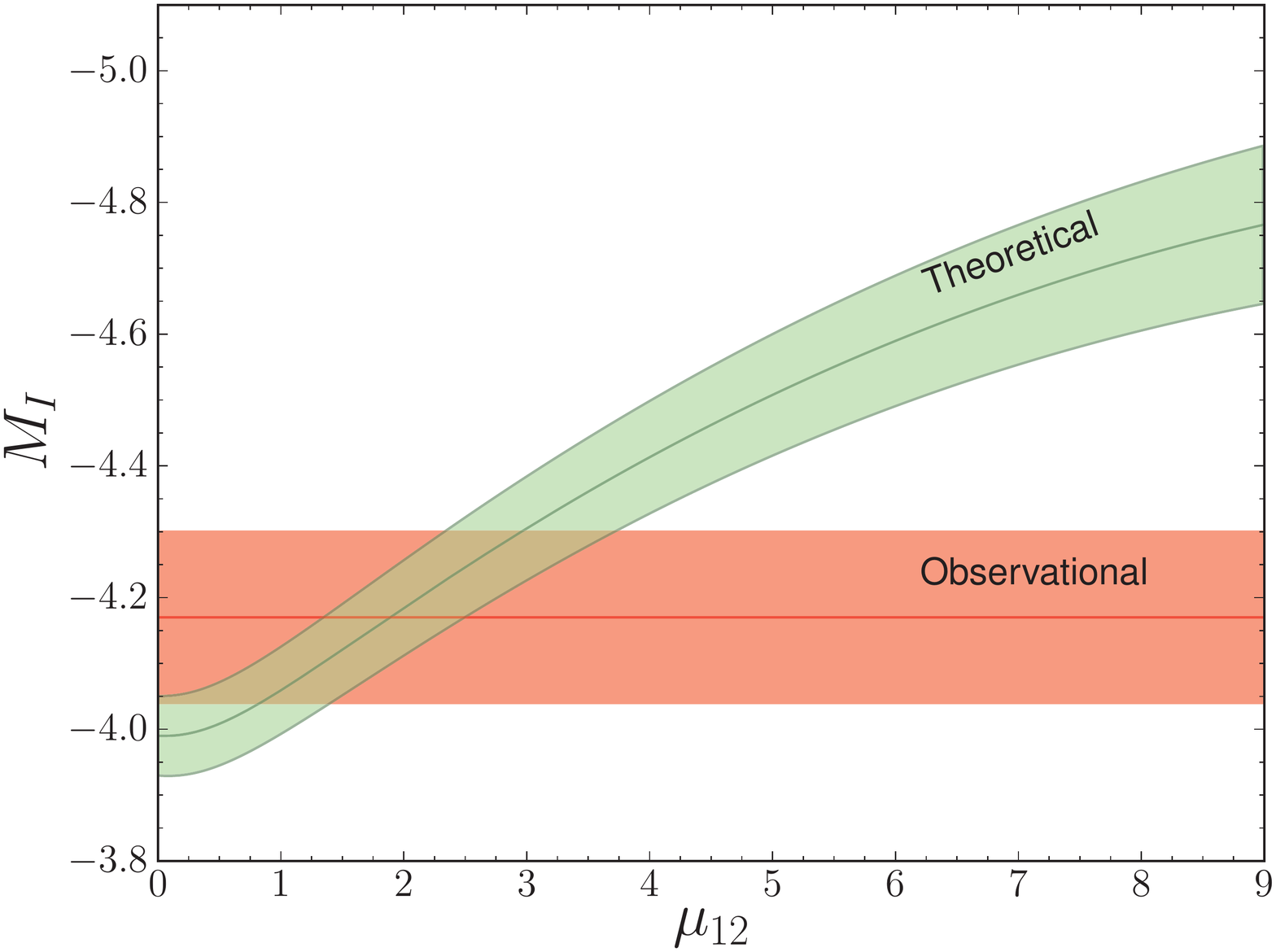}
\includegraphics[width=8cm]{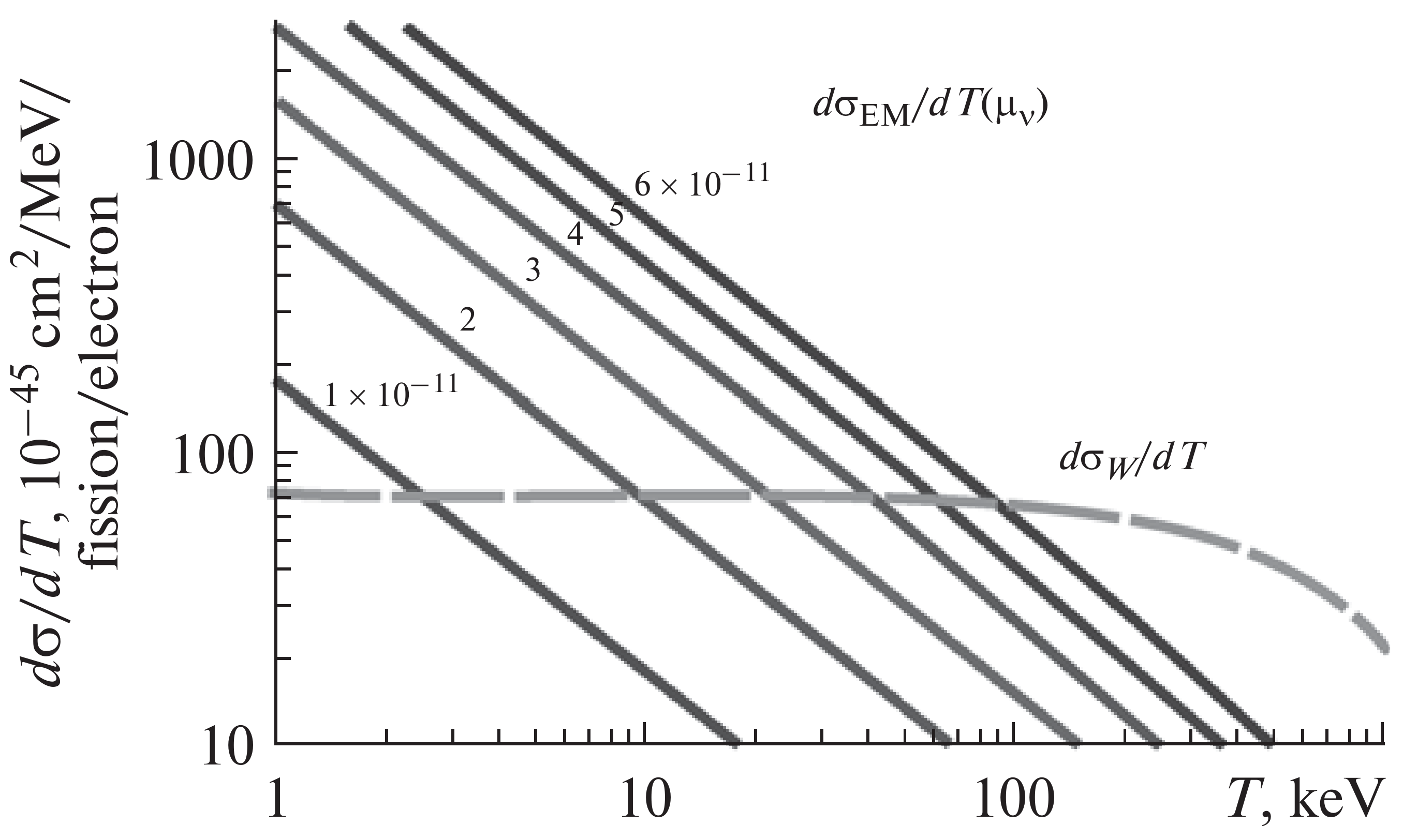}
\caption{Left: Green band: calculated brightness (magnitude) of stars in the tip of the red-giant branch (tRGB) as a function of $\mu_{12}=\mu_{\nu}\times10^{-12}~\mu_{\rm B}$. Red band: measured brightness of tRGB stars in cluster M50 . Reprinted with permission from \cite{Viaux2013b}. Right: Differential cross-section for electron-neutrino scattering as a function of the electron recoil energy $T$. Electromagnetic and weak contributions are shown. Reprinted with permission from \cite{Beda2013}.}
\label{fig:neutrinos}
\end{figure}
There are a number of approaches for setting limits on $\mu_{\nu}$. Some of the best measurements to date come from astronomical observations of the red-giant branch in globular clusters \cite{Diaz2015,Viaux2013}. The evolution of these stars is sufficiently well understood that they act as an indirect probe for the presence of $\mu_{\nu}$; a sizable magnetic moment would enhance neutrino production via plasmon decay, thus increasing the core mass and brightness of the tip of the red-giant branch (tRGB) \cite{Bernstein1963,Raffelt1990}. The green band in the left-hand plot of Fig.~\ref{fig:neutrinos} shows the predicted brightness (magnitude) $M_I$ of the tRGB as a function of $\mu_{\nu}$. The red band represents the observed brightness, displaying good agreement with $\mu_{\nu}=0$ and giving the limit
\begin{equation}
\mu_{\nu}\leq4.5\times10^{-12}~\mu_{\rm B}~(95\%~{\rm C.L.}).
\end{equation}
The largest source of error in the experiment is from the uncertainty in the distance of the stars; significant improvement is expected from the upcoming GAIA mission.

The Gemma experiment presents an alternative approach to measuring $\mu_{\nu}$ by examining electron-neutrino scattering in a reactor \cite{Beda2013}. The differential cross-section with respect to the electron energy, $d\sigma/dT$, would have an electromagnetic contribution in the presence of $\mu_{\nu}$. The right-hand plot of Fig.~\ref{fig:neutrinos} shows the calculated dependence of this differential cross-section on $T$ for both the weak and electromagnetic components --- at low $T$ the weak contribution becomes almost constant. By measuring the differential cross-section at low $T$ the following limit was placed:
\begin{equation}
\mu_{\nu}\leq2.9\times10^{-11}~\mu_{\rm B}~(90\%~{\rm C.L.}).
\end{equation}
Work is currently under way to prepare the next generation of the experiment, with a predicted sensitivity of $1\times10^{-11}~\mu_{\rm B}$.

Due to the near-massless nature of neutrinos, and hence their ultra-relativistic motion under most circumstances, limits on the magnetic dipole moment can be restated identically as limits on the electric dipole moment \cite{Okun1986}, assuming instead $\mu_{\nu}=0$.

\section{Electric Dipole Moments}

\subsection{Electron}
As the most stable lepton, the electron has the most precisely measured EDM. The current best limit on $d_e$ is that of the ACME collaboration \cite{Baron2014}\footnote{The value quoted here uses an average of the two calculated values of the effective electric field, $\mathcal{E}_{\rm eff}$ in ThO \cite{Fleig2014,Skripnikov2015}.}:
\begin{equation}
d_e<9.3\times10^{-29}~e\cdot{\rm cm}=4.8\times10^{-18}~e\hbar/2m_ec~(90\%~{\rm C.L.}),
\end{equation}
which improved on the previous limit by an order of magnitude \cite{Hudson2011}. Using Equation~\ref{eq:NS1} we find that this measurement probes one-loop NP effects at the ${\sim}10$~TeV mass scale if the $CP$-violating phases associated with the NP are of order 1.

The sensitivity of the ACME experiment was afforded by measuring the precession of the electron inside a thorium monoxide (ThO) molecule, providing large values of $\mathcal{C}$, $\tau$, $\mathcal{E}_{\rm eff}$ and $\dot{N}$ (cf. Equation~\ref{eq:statsens}). In particular, ThO has a huge effective internal electric field $\mathcal{E}_{\rm eff}\approx80~{\rm GV/cm}$ \cite{Fleig2014,Skripnikov2015}, representing around a $10^6$ enhancement over typically achievable laboratory fields.

\begin{figure}[!ht]
\centering
\includegraphics[width=11cm]{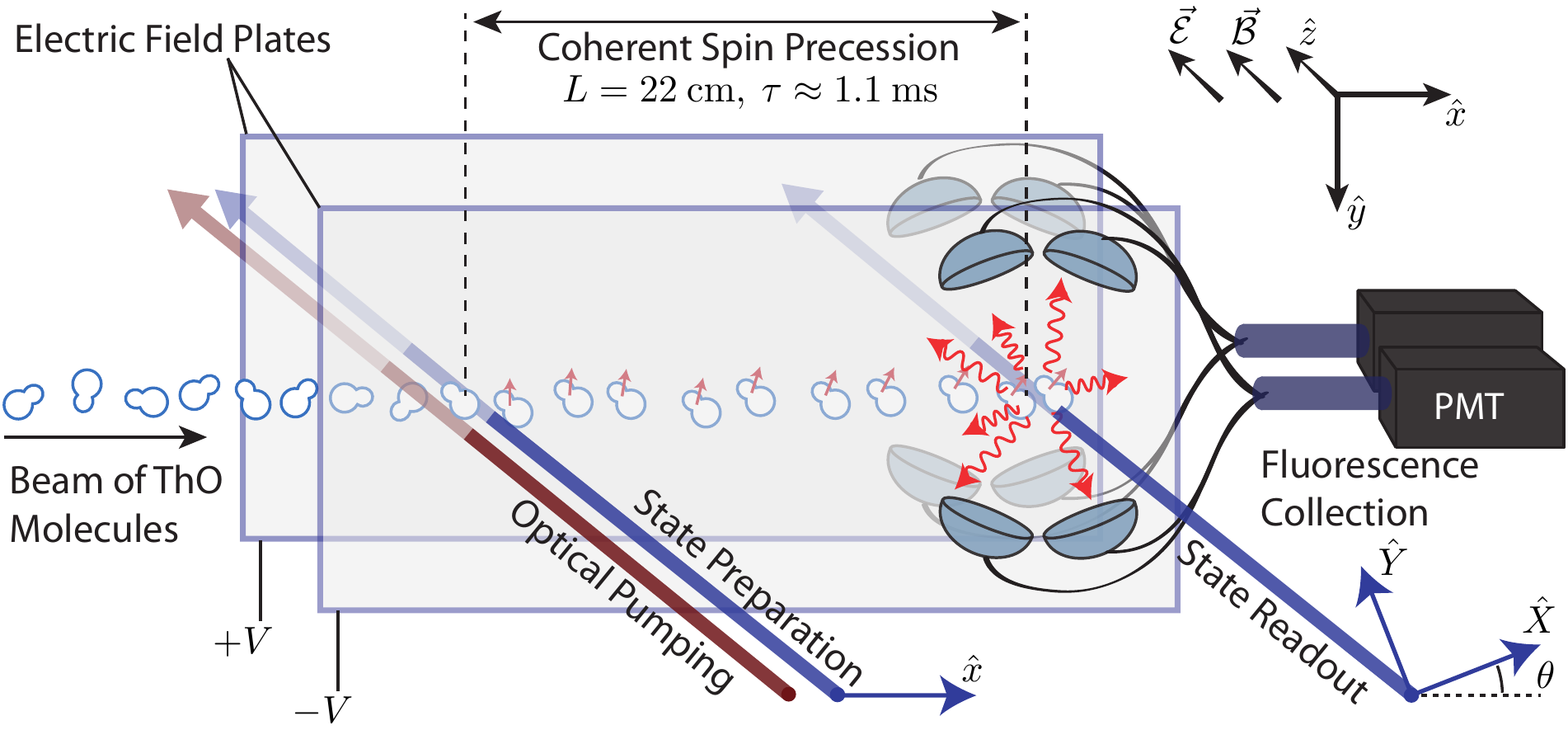}
\caption{Schematic of the ACME experiment to measure $d_e$. See main text for description.}
\label{fig:edm_expt}
\end{figure}

The quantum mechanical phase associated with the time-integrated interaction energy, $\phi=d_e\mathcal{E}_{\rm eff}\tau/\hbar$, is equivalent to a precession angle of the electron's spin. The experimental scheme amounts to measuring the accumulated precession angle of an electron in ThO in the presence of electric and magnetic fields. A schematic of the apparatus is shown in Fig.~\ref{fig:edm_expt}.
ThO molecules are produced by a cryogenic buffer gas beam source \cite{Hutzler2011} and undergo rotational cooling before entering a magnetically shielded interaction region with uniform applied electric and magnetic fields. The molecules are optically pumped from the ground electronic state to the EDM-sensitive $H$-state. This is a triplet spin state and a superposition of the form $\left|M=-1\right\rangle\pm\left|M=+1\right\rangle$, where $M$ is the total angular momentum projection onto the quantisation axis, is prepared via laser excitation. The molecules then traverse 22~cm in $\tau{\sim}1$~ms, accumulating a phase $\phi$. The state evolves to $e^{-i\phi/2}\left|M=-1\right\rangle\pm e^{+i\phi/2}\left|M=+1\right\rangle$. This phase is then read out by re-projecting back onto orthogonal spin states via laser excitation and observing the resulting fluorescence. The accumulated phase is given by
\begin{equation}
\phi=-2d_e\hat{S}\cdot\vec{\mathcal{E}}_{\rm eff}\tau-2\mu_e\hat{S}\cdot\vec{\mathcal{B}}\tau+\cdots=\phi_{\rm EDM}+\phi_{\rm MDM}+\cdots
\end{equation}
where $\vec{\mathcal{B}}$ is the applied magnetic field. Because the EDM-associated phase, $\phi_{\rm EDM}$, is dwarfed by other phase components, it is vital to effectively isolate it. This is achieved through experimental switches. For example, repeating the experiment after a reversal of the laboratory field that orients the molecule yields
\begin{equation}
\phi(-\mathcal{E})-\phi(+\mathcal{E})=2d_e\mathcal{E}_{\rm eff}\tau/\hbar+\cdots;
\end{equation}
the precession due to the interaction with the magnetic field has been removed (save for possible systematic effects). For more information on experimental details and systematics the interested reader is directed to \cite{Hutzler2014,Spaun2014,Hess2014}.

\begin{figure}
\centering
\includegraphics[width=7cm]{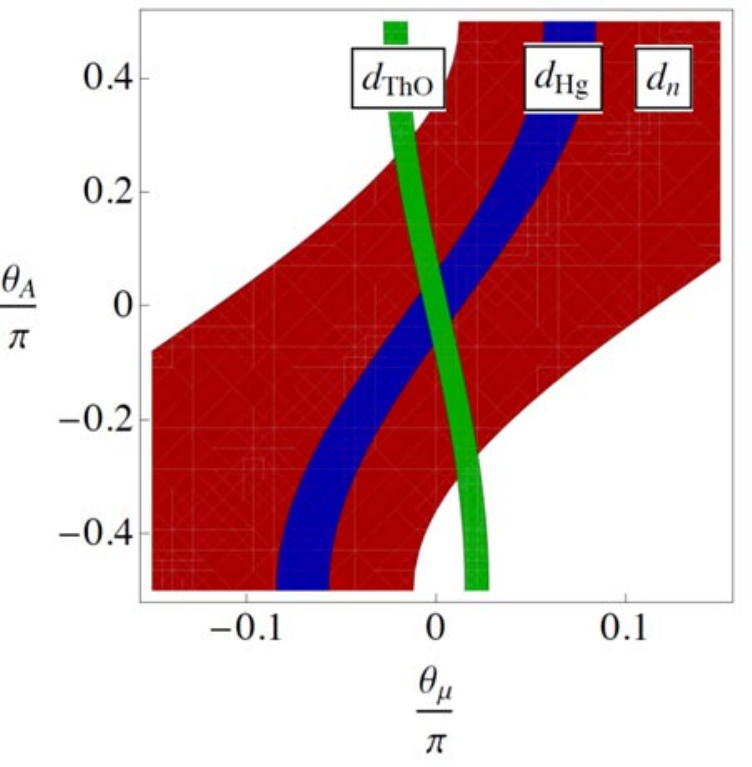}
\caption{Constraints on MSSM $CP$-violating phases from EDM measurements. $d_{\rm ThO}$, $d_{\rm Hg}$ and $d_n$ are the measured EDMs of thorium monoxide, mercury and the neutron respectively. First-generation superpartners are assumed to have mass 2~TeV. Reprinted with permission from \cite{Dall2013,Ritzprivate}.}
\label{fig:edm_phases}
\end{figure}

The limit on $d_e$ from ACME has resulted in strong constraint of many BSM theories. In particular, many SUSY models make predictions of $d_e$ around the range of the current limit --- consistency with the experimental limit is found through tuning of model parameters. This is illustrated in Fig.~\ref{fig:edm_phases} (updated version of \cite{Dall2013}).
The figure shows schematically the parameter space of the minimal supersymmetric model (MSSM) that is constrained by various EDM limits. $\theta_A$ and $\theta_{\mu}$ are two $CP$-violating phases. $d_{\rm ThO}$, $d_{\rm Hg}$ and $d_n$ correspond to the EDM measurements in ThO, the mercury atom and the neutron respectively. The plot assumes that the mass of first-generation superpartners is 2~TeV, a lower limit consistent with direct searches at the LHC. We see that the parameter space of the $CP$-violating phases in the MSSM is strongly constrained by this set of measurements, with the tightest bounds provided by the ACME experiment. The lack of evidence for $CP$-violation in EDMs leads to the `SUSY $CP$ problem', which can be solved either by allowing `unnaturally small' phases or large superpartner masses \cite{Hamed2005}. The latter in turn would suggest further fine tuning to avoid the Higgs renormalization problem and hence has motivated investigation of alternative solutions to the SUSY $CP$ problem \cite{Schmaltz2000,Cruz2005,Ishiduki2009}.

The ACME experiment is currently being upgraded to make an improved measurement of $d_e$. The upgrades will include a higher flux beam source, optimised beam geometry, improved molecule state preparation and improved fluorescence collection/detection. It is anticipated that the statistical sensitivity will be increased by another order of magnitude. A successful measurement at such a sensitivity would probe NP at the 10s of TeV scale.

As mentioned in Sec.~\ref{sec:mugm2} it is useful to compare the limit on $d_e$ to the discrepancy on $a_{\mu}$. From \cite{Giudice2012} we have, assuming na\"{i}ve scaling:
\begin{equation}
d_e^{\rm NS}\approx1\times10^{-11}~\Delta a_e\tan\phi_e~e\cdot{\rm cm}.
\end{equation}
Using Equation~\ref{eq:et_gm2_disc} we then have
\begin{equation}
d_e^{\rm NS}\approx6\times10^{-25}~e\cdot{\rm cm}=3\times10^{-14}~e\hbar/2m_ec
\end{equation}
where we have assumed that the $CP$-violating phase $\tan{\phi_e}=1$. The current limit on $d_e$ is over three orders of magnitude lower than this value, potentially constraining $\tan\phi_e<10^{-3}$ or suggesting a deviation from na\"{i}ve scaling.

\subsection{Muon}
The current limit on the EDM of the muon was measured by the same experiment that measured $a_{\mu}$ \cite{Bennett2009}:
\begin{equation}
d_{\mu}\leq1.8\times10^{-19}~e\cdot{\rm cm}=1.9\times10^{-6}~e\hbar/2m_{\mu}c~(95\%~{\rm C.L.}).
\end{equation}
This value is extracted from the measurement of $a_{\mu}$ by considering an additional contribution to the anomalous frequency (cf. Eq.~\ref{eq:mu_wa}):
\begin{equation}
\vec{\omega}_a=a_{\mu}\frac{e\vec{B}}{mc}-2d_{\mu}/\hbar\left(\vec{\beta}\times\vec{B}+\vec{E}\right).
\end{equation}
$\vec{B}$ is the magnetic field applied to produce the desired cyclotron motion and $\vec{E}$ is the electric field from electrostatic focussing elements. The former produces a motional electric field which is in fact much larger than the latter, and hence the latter term is generally ignored. 

Recall that the spin vector precesses about the direction of $\vec{B}$ (vertical); i.e. the precession is in the plane of the storage ring. Additional precession due to $d_{\mu}$ is about the vector ($\vec{\beta}\times\vec{B}$). Adding these vectors together, the spin is thus tipped out of the plane with maximum (zero) vertical excursion when the spin is directed approximately radially (azimuthally). The presence of $d_{\mu}$ would also slightly increase the spin precession frequency. 

\begin{figure}[!ht]
\centering
\includegraphics[trim={0 14.25cm 0 0},clip,height=5cm]{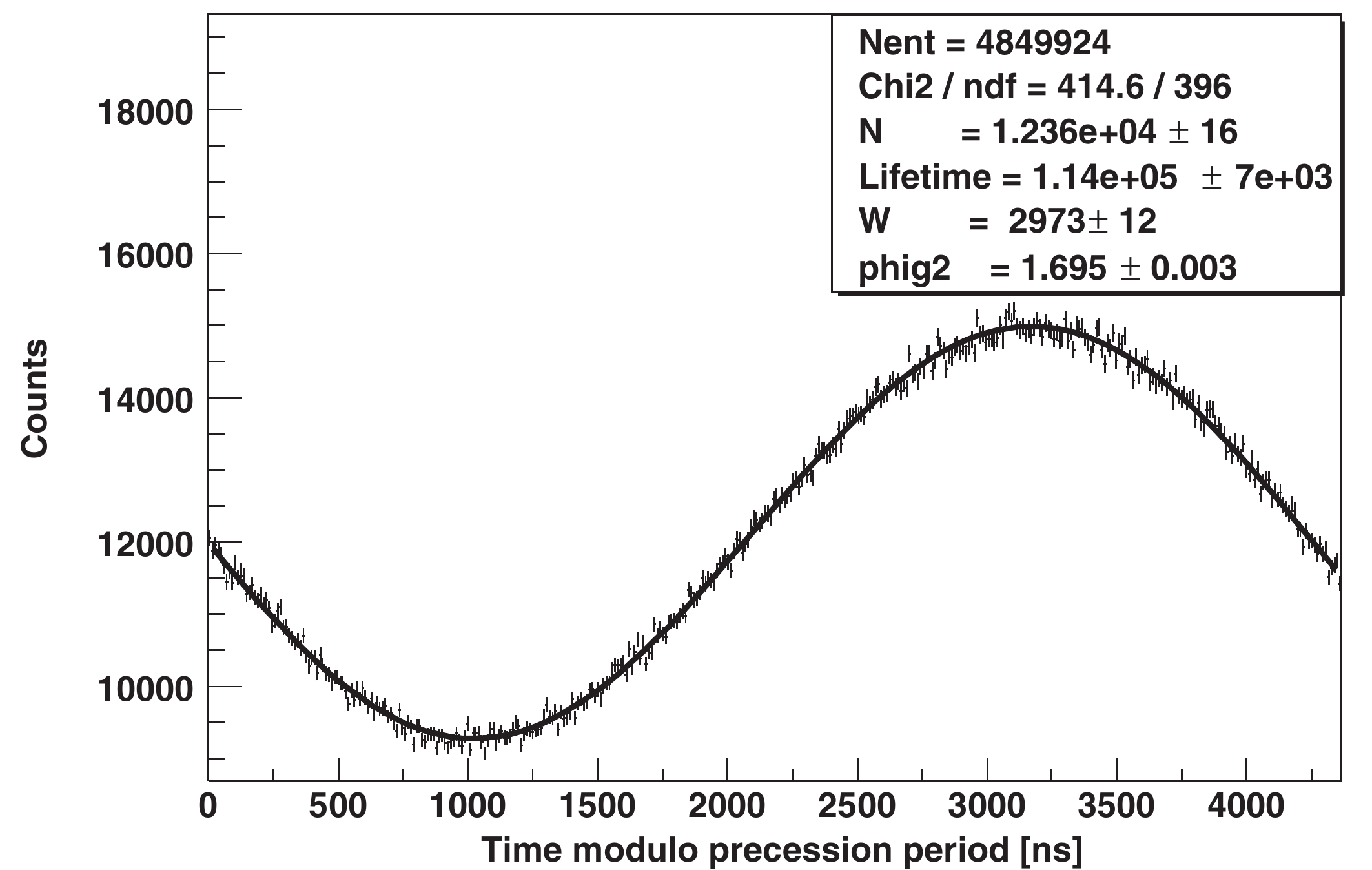}
\includegraphics[trim={0 14.05cm 0 0},clip,height=5cm]{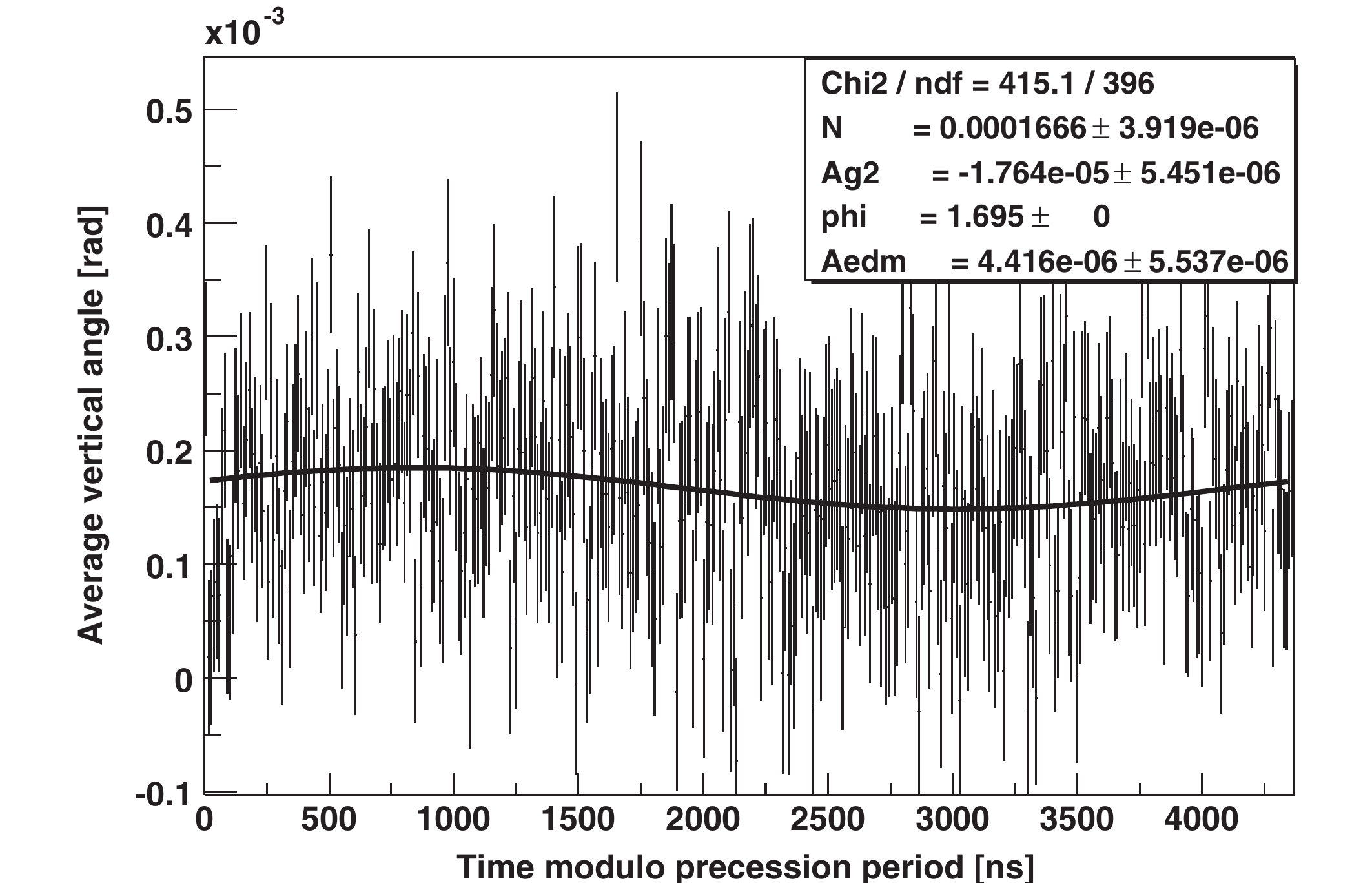}
\caption{Example data from E821 used to extract $d_{\mu}$. Left: Variation in electron count rate. These data are used to determine the anomalous precession frequency and phase. Right: Average vertical angle of the muon spin. The fit is a two component sinusoid of frequency $\omega_a$, in phase and $\pi/2$ out of phase. Reprinted with permission from \cite{Bennett2009}.}
\label{fig:muedm_data}
\end{figure}

The experiment preferentially selects high-energy electrons from the muon decay which are correlated with the muon spin direction, and, due to the muon's motion, move approximately azimuthally in the lab frame. Due to this selectivity, the observed vertical oscillation due to $d_{\mu}$ is $\pi/2$ out of phase with the signal size oscillation (due to $a_{\mu}$). This helps to distinguish effects due to $d_{\mu}$ from $a_{\mu}$-correlated systematic effects. Example data used to extract $d_{\mu}$ are shown in Fig.~\ref{fig:muedm_data}.
The left-hand plots shows the variation in overal signal size, modulo the precession period. Vertical oscillation of the spin is observed through the use of an array of scintillators and drift chambers. The precession frequency and phase extracted from the overall signal is used to fit the vertical oscillation. Averaging over the available data produces an amplitude of vertical oscillation, and hence $d_{\mu}$, which is consistent with zero at a level of less than $1\sigma$.

The null measurement of $d_{\mu}$ helps to bolster the $a_{\mu}$ discrepancy --- a modification of $|\vec{\omega}_a|$ due to $d_{\mu}$ that would account for the discrepancy is now unlikely \cite{Bennett2009}. If the discrepancy were instead caused by NP, na\"{i}ve scaling would imply \cite{Giudice2012}
\begin{equation}
d^{\rm NS}_{\mu}\approx1.8\times10^{-22}~e\cdot{\rm cm}=1.9\times10^{-9}~e\hbar/2m_{\mu}c,
\end{equation}
which is significantly lower than the current limit. Under this assumption the measurement of $d_{\mu}$ is a less stringent test of NP; however, other models offer alternative scalings (e.g. \cite{Ellis2002,Ibrahim2001}) and $d_{\mu}$ remains a complementary test of such models. An alternative approach to the measurement of $d_{\mu}$ could provide significantly increased sensitivity, at around the $10^{-24}~e\cdot{\rm cm}$ level \cite{Semertzidis2001}. Such an improvement would allow for direct comparison with the prediction above.

\subsection{Tau}
As with $a_{\tau}$, the relative difficulty of measuring the electromagnetic form factor of taus prevents utilising the sensitivity to NP afforded by its larger mass. The current limit on the electric dipole form factor, $d_{\tau}$, is also set through a comparison with QED theory; measurement of the spin density matrix for the process $e^+e^-\rightarrow\gamma^*\rightarrow\tau^+\tau^-$ by the Belle collaboration gives \cite{Belle2003}
\begin{align}
-2.2\times10^{-17}<{\rm Re}(d_{\tau})&<4.5\times10^{-17}~e\cdot{\rm cm},\\
-2.5\times10^{-17}<{\rm Im}(d_{\tau})&<0.8\times10^{-17}~e\cdot{\rm cm}.
\end{align}
This is equivalent to
\begin{align}
-3.8\times10^{-3}<{\rm Re}(d_{\tau})&<8.0\times10^{-3}~e\hbar/2m_{\tau}c,\\
-4.5\times10^{-3}<{\rm Im}(d_{\tau})&<1.4\times10^{-3}~e\hbar/2m_{\tau}c.
\end{align}
We note that a similar limit can, with some assumptions, be inferred through a consideration of the contribution of $d_{\tau}$ to $d_e$ \cite{Grozin2009}.

As with $a_{\tau}$, significant increase in precision is anticipated with future experiments; the 40-fold increase in luminosity at SuperKEKB \cite{SuperKEKB} will give Belle II a roughly six-fold increase in statistical sensitivity, and there are proposals for measuring $d_{\tau}$ with polarised electron beams that could provide two orders of magnitude greater sensitivity \cite{Bernabeu2007}.

\section{Outlook and Conclusion}
Recent progress in the measurement and calculation of lepton dipole moments continues to provide some of the most stringent tests of the SM and sensitive searches for NP, complementary to the ongoing work of colliders such as the LHC. We see that the greatest precision is achieved with measurements in the electron: the precision with which $\mu_e$ has been measured is around $10^{-14}~e\hbar/2m_ec$, whilst the limit on $d_e$ is at the $10^{-18}~e\hbar/2m_ec$ level, the latter probing a significantly higher mass scale. However perhaps the most tantalizing current datum is the discrepancy between experiment and theory for $a_{\mu}$. While the 3$\sigma$ disagreement may be hinting at NP, the absence of an analogous inconsistency in $d_e$ lends little support under a na\"{i}ve scaling model --- though $d_e$ requires $CP$-violation in the NP and $a_e$ does not. While the planned more precise measurement of $a_{\mu}$ should help resolve the issue, future measurements of both $a_e$ and $d_{\mu}$ will also be important in broadening our view of this frontier. Meanwhile, the measurement of $a_e$ continues to provide us with our most precise test of QED and the most precise determination of $\alpha$, and the measurement of $d_e$ has allowed us to probe NP at mass scales in the 10~TeV range. Whilst there is not yet clear evidence for BSM physics, all measurements continue to refine the parameter space within which our best theories should operate, and the continued evolution of the field allows us to remain hopeful that the next important discovery could be close by.

\subsection*{Acknowledgments}
The author thanks the organisers for the invitation to this conference, David DeMille for much helpful advice in preparing this manuscript, and all who contributed to the ACME collaboration.

\end{document}